# Document Retrieval using Predication Similarity

Kalpa Gunaratna[1]

Kno.e.sis Center, Wright State University, Dayton, OH 45435 USA
`kalpa@knoesis.org`

**Abstract.** Document retrieval has been an important research problem over many years in the information retrieval community. State-of-the-art techniques utilize various methods in matching documents to a given document including keywords, phrases, and annotations. In this paper, we propose a new approach for document retrieval that utilizes predications (subject-predicate-object triples) extracted from the documents. We represent documents as sets of predications. We measure the similarity between predications to compute the similarity between documents. Our approach utilizes the hierarchical information available in ontologies in computing concept-concept similarity, making the approach flexible. Predication-based document similarity is more precise and forms the basis for a semantically aware document retrieval system. We show that the approach is competitive with an existing state-of-the-art related document retrieval technique in the biomedical domain.

## 1  Introduction

Finding related documents is an interesting research problem in text and document retrieval. Keyword co-occurrence, matching combination of keywords, and cosine similarity of term vectors are some of the techniques used to match documents. In the simplest form, documents can be indexed using keywords and these keyword indices can be used to retrieve related documents, but this does not handle semantic similarity between documents. By semantic similarity, we mean a matching that goes beyond lexical similarity computations like exact matching of keywords. Furthermore, keyword-based systems including advanced systems like PubMed can handle more than one keyword search query using keyword co-occurrence but we are interested in retrieval based on triples and not just concepts (i.e., keywords).

A semantically aware documents retrieval system can help a typical user who needs to get related documents even when he is not completely sure of exactly what keywords or phrases to use for the search. Furthermore, if related documents can be

---

fetched utilizing triples (i.e., semantic predications) which are extracted from documents, it can provide more precise and also semantically matched results. For example, a query to find documents that contain triples like "ASPIRIN TREATS HEADACHE" is expected to retrieve documents that have drugs related to ASPIRIN treating diseases like HEADACHE. This is different from a keyword-based search query constructed by concatenating keywords where order of keywords does not matter and no semantic matching is performed.

We propose a document retrieval technique based on semantic matching of triples (predications) extracted from documents in the biomedical domain.

The Biomedical Knowledge Repository (BKR) [6] is a repository of integrated biomedical data from literature, structured databases, and terminological knowledge sources like Unified Medical Language System (UMLS) [7]. BKR represents the integrated information in RDF, for example, the RDF triple, which is also called a predication, "`lipoprotein`→`affects`→`inflammatory_cells`" was extracted by the text mining tool SemRep [1] from a MEDLINE[2] journal article (with PubMed identifier PMID: 17209178). It states that `lipoprotein` (denoted as "subject" of the RDF triple) `affects` (denoted as "property" of the triple) `inflammatory_cells` (denoted as the "object" of the triple). Each document can be represented using a set of extracted predications like these. In this approach, we compute the similarity between sets of predications to derive the similarity between documents. The proposed approach contributes in the following ways to document retrieval:

1) More precise – it searches at the predication level rather than words.
2) More flexible – it uses semantic similarity and hence covers more document matches than pure lexical similarity matches.
3) Semantically aware – it takes into consideration the "context" (captured by predications) in which the user searches related documents using sentences.

The rest of the paper is organized as follows. Section 2 presents background and related work in the area and Section 3 presents the details of our proposed approach. Section 4 brings preliminary results and Section 5 discusses advantages and limitations of our approach. We conclude with future work in Section 6.

## 2 Background and Related Work

Concept similarity computation is a popular topic in the biomedical research community. There exist several kinds of similarity measures in the literature that use distance between concepts (number of nodes/ edge counting), node features (e.g., proportion of shared ancestors), and information content (e.g., frequency of a concept in a given corpus) [3]. In this paper, we consider using a simple, yet powerful measurement to capture ontological similarity of concept pairs. The use of a simple similarity measure is needed because of the large number of similarity pair computations re-quired in our application. Batet et. al [3] proposed a measure of dissimilarity based on the shared number of ancestors between two concepts. However, its value is not normalized, making it unsuitable for our application.

---

[2] MEDLINE contains journal citations and abstracts for biomedical literature from around the world. PubMed is the interface to MEDLINE.

There are other systems that try to index and retrieve related predications. Cohen et. al [4] proposed an indexing mechanism for predications and a retrieval mechanism. Even though it has advanced retrieval capabilities like leaving part of the predication empty, it has no flexibility in matching related predications (i.e., no semantic similarity). TripleRank [5] is an authoritative ranking mechanism for triples based on the "popularity" of triples. It is related to our system as it ranks triples in a given context, but does not consider similarity or relatedness between triples.

Our proposed approach is different from indexing systems and keyword based retrieval mechanisms as it consists of a flexible semantic matching component.

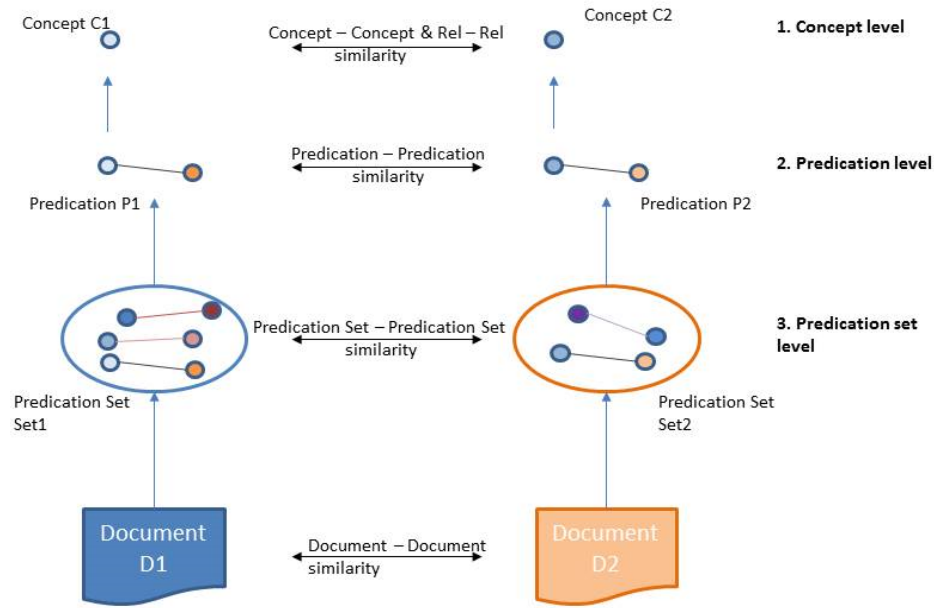

**Fig. 1.** Overview of the method

## 3      Approach

We are interested in computing similarity between documents using predications. A document can be represented as a set of predications. Furthermore, similarity between two sets of predications belonging to two documents can be used to compute the similarity between the two documents. Fig. 1 shows how document – document similarity computation is decomposed into three stages of similarity computation.
1. Compute concept - concept and relationship-relationship similarity.
2. Compute predication - predication similarity.
3. Compute predication-set – predication-set similarity.

Predications in the BKR have concept (class) instances as subjects and objects. We are interested in finding out concept level similarity for predications and hence, we represent each subject and object of predications with its assigned concept.

### 3.1 Concept – Concept similarity

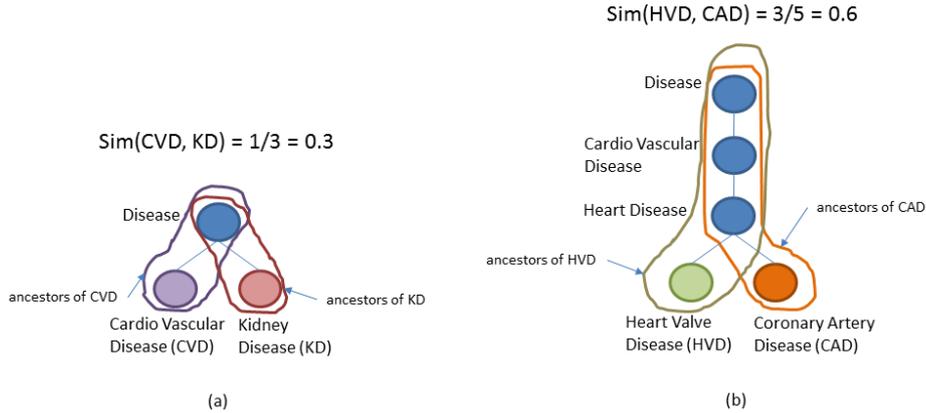

**Fig. 2.** Jaccard similarity computation examples.

Since there are many predications in the BKR, we try to use a simple similarity measure for concept – concept similarity. The idea is to use the proportion of shared ancestors between two concepts as a measure of their similarity. We leverage hierarchical relations in the UMLS Metathesaurus, a terminology integration system, to compute the set of ancestors for each concept. We use the Jaccard coefficient to quantify the overlap between two sets of ancestors. The similarity can be computed as shown in equation 1. Of note, in this similarity computation, we add the concept itself to its set of ancestors in order to preserve high similarity for concepts that appear lower in the hierarchy. Fig. 2(a) and 2(b) show the behavior of Jaccard similarity computation for concepts. When concepts are higher in the concept hierarchy, they have very abstract meaning and hence, there similarity is expected to be lower (Fig. 2(a)) than the ones that appear lower in the hierarchy where they have very specific meaning (Fig. 2(b)). The Jaccard similarity value computed in this way varies between 0 and 1. Note that we follow a similar procedure for relationship – relationship similarity computation by using the hierarchy of relationships from the UMLS Semantic Network.

$$Jaccard(C1, C2) = \frac{\#\ shared\ ancestors\ between\ C1\ and\ C2}{\#\ total\ concepts\ in\ C1\ and\ C2} \quad (1)$$

### 3.2 Predication – Predication similarity

We compute the similarity between two predications as the average pairwise similarity of subject, predicate, and objects pairs. Similarity between a predication P1 and predication P2 denoted as Sim(P1, P2) is computed as shown in equation 2. *Ws, Wr,*

*and Wo* are weights associated with similarity values between subjects, predicates, and objects, respectively. Since we get the average similarity over subject, predicate, and object pairs and each similarity values is between 0 and 1, Sim(P1, P2) is always between 0 and 1.

$$Sim(P1, P2) = \frac{Ws * Sim(S1, S2) + Wr * Sim(R1, R2) + Wo * Sim(O1, O2)}{Ws + Wr + Wo} \quad (2)$$

An example similarity computation of two predications is shown in Fig. 3. In the example, all weights are equal to 1.

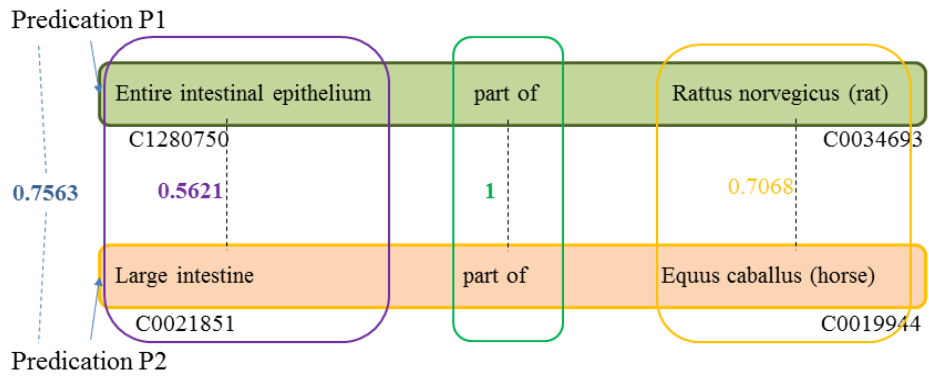

**Fig. 3.** predication – predication similarity computation

### 3.3 Predication-set – Predication-set similarity

Similarity between sets of predications is computed according to the formula shown in equation 3. This is an accepted technique to compute the similarity between two sets [2] based on the similarity between their members. The intuition is to compute predication – predication similarity between the two sets in both directions picking the maximum value for each predication and taking the average across the total number of predications.

$$Sim(Set\ S1, Set\ S2) = \frac{1}{m+n} * \left( \sum_k \max_p(Sim(Pk, Pp)) + \sum_p \max_k(Sim(Pk, Pp)) \right) \quad (3)$$

## 4 Experiment

Our main contribution in this proposed approach is computing predication - predication similarity utilizing concept-concept similarity. To evaluate this novel idea, we take related document retrieval as a use case. PubMed related citation search provides reference related citations for a given PubMed article and we evaluate our related document retrieval results against this gold standard.

### 4.1 Evaluation setting

For our evaluation, we selected a subset of articles from MEDLINE and retrieved their predications from the 2013AB version of the BKR. Our evaluation is limited in size and only serves as a proof of concept for our proposed approach. The document sample for the evaluation is selected as follows. First, we randomly selected 30 documents from MEDLINE citations. Then, for each of these documents, we retrieved the top 30 related citations from the PubMed related citation search. Our document sample includes the 30 seed documents, as well as the top 30 documents retrieved for each of them, resulting in 907 documents after removing duplicates.

### 4.2 Implementation details

The prototype is developed using the Java programming language and we used the Virtuoso[3] triple store to store predications. Similarity values for concept and relationship pairs, predication pairs, and document pairs are stored in memory as key-value pairs using BerkeleyDB[4] database for rapid access.

---

[3] http://virtuoso.openlinksw.com/

[4] http://www.oracle.com/technetwork/database/databasetechnologies/berkeleydb/overview/index.html

### 4.3 Results

We measure precision, recall, and F-measure against the PubMed related citation gold standard. Computation of precision, recall, and F-measure are defined in equations 4, 5, and 6, respectively.

$$precision = \frac{\#\ releveant\ documents\ \cap\ \#\ retrieved\ documents}{\#\ retrieved\ documents} \quad (4)$$

$$recall = \frac{\#\ relevant\ documents\ \cap\ \#\ retrieved\ documents}{\#\ relevant\ documents} \quad (5)$$

$$F - measure = 2 * \frac{precision * recall}{precision + recall} \quad (6)$$

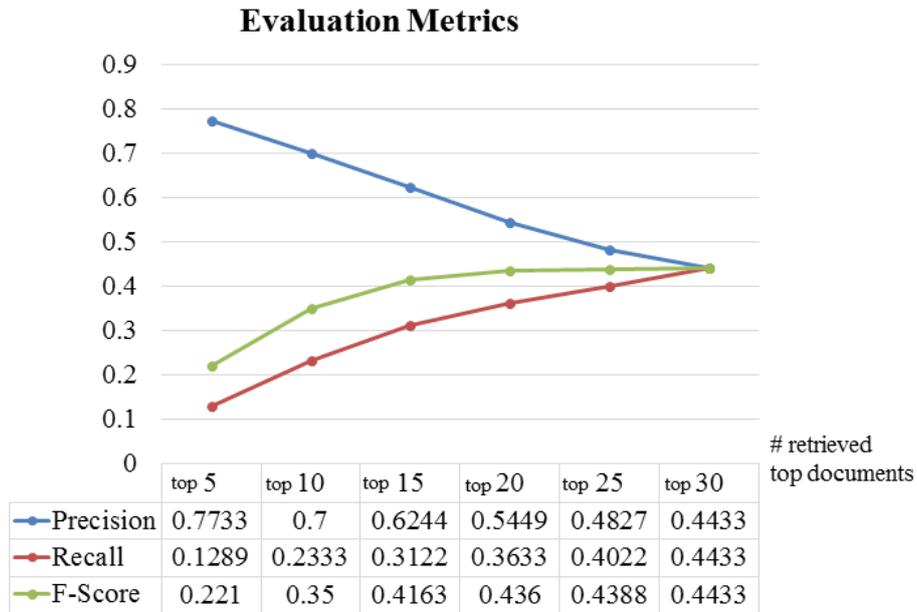

**Fig. 4.** Results against the PubMed related citation gold standard using top n documents

The preliminary results against the PubMed related citation gold standard is shown in Fig. 4. We measured precision, recall, and F-measure for the top n documents as shown in Fig. 4. For the top 5 documents, 77% of the documents retrieved by our approach are relevant, demonstrating that it can retrieve documents with high precision. The reason for lower recall is that we did not adjust weights in the predication – predication similarity computation and prune similarity values that only matched part of the predications. Matching only part of the predications does not mean they are similar as they can be out of context.

## 5    Discussion

Our results showed promise for a new way of computing document similarity using semantic predications. Moreover, we can further improve the precision and recall by following the ideas below.

- Improving precision: We haven't filtered out predication pairs that have very low similarity values. This artificially lowers the document - document similarity ranking. Introducing a threshold to filter out low-similarity predication pairs can improve precision.
- Improving recall: We haven't experimented with suitable weights for subject, predicate, and object pairs in predication – predication similarity computation. Learning suitable weights for the related document retrieval use case can improve recall as it removes noise in predication similarity computation.

### 5.1    Advantages

Our bag-of-predications approach has clear advantage over bag-of-words document retrieval systems as it is semantically-aware. It makes use of hierarchical relationships between concepts, as well as hierarchies of relationships, to measure concept similarity. Concept similarity adds flexibility in retrieving related documents, whereas bag-of-words approaches can only leverage word occurrence or co-occurrence (i.e., utilizing probabilities). Because it is based on predications, not keywords, our approach can also provide more precise results, as it captures the context of the query. For example, the predication "*ASPIRIN TREATS HEADACHE*" is not simply the concatenation of the three keywords "*ASPIRIN*", "*TREATS*", and "*HEADACHE*". It expresses the precise treatment relation between the drug and the disease. For this reason, our approach is more precise than traditional document retrieval models.

Furthermore, we could also use predication – predication similarity to provide question answering or exploration capabilities to a user. For example, a user can ask a question like "give me related predications to *ASPIRIN TREATS HEADACHE*" or "find *<what?> TREATS HEADACHE*". In the first example, the user can explore related predications and in the second example, he can find out what drugs (or interventions) can treat headache.

### 5.2    Limitations

There are three different limitations with the current prototype.

1. Limitations with SemRep
   (a) SemRep uses a template-based predication extraction and hence it can miss extraction of some predications from articles.

(b) Because SemRep does not handle co-reference resolution of named entities, it cannot extract predication across sentences and hence it can miss extraction of predications that span across sentences.
2. Limitations with similarity
  (a) Concept – concept similarity needs to be evaluated in UMLS. The simple notion of shared proportion of ancestors between two concepts (measured with the Jaccard coefficient) has never been evaluated on UMLS hierarchies, even though it has been tested in other domain datasets like genes.
  (b) Weights of the predication – predication similarity needs to be calibrated for better performance. As of now, we use the simplest representation of weights, with all weights equal to 1.
  (c) The scale of the current evaluation is small. Moreover, our evaluation is limited in scope (i.e., against PubMed related citation search results.)
3. The current prototype cannot handle the large number of similarity computations required to scale to the whole MEDLINE corpus.

# 6 Future Work and Conclusion

## 6.1 Future Work

We plan to address the limitations mentioned in the previous section except the limitations due to SemRep as improvements to SemRep are out of our scope. We plan to evaluate our concept – concept similarity metric on the UMLS, and to adjust and learn weights empirically for predication similarity. Also a large-scale evaluation with an independent test dataset (i.e., other than PubMed related citations) needs to be conducted. We also would like to address the scalability issues related to computation by adapting cluster-based computations and use efficient data storage for large indices. As mentioned in Section 5, it is worthwhile to investigate how to further improve precision and recall by adapting a threshold variable and weights adjustment.

## 6.2 Conclusion

In this paper, we proposed a document retrieval approach that leverages semantic predications (in other terms, triples) extracted from these documents. We introduced the idea of using a bag-of-predications approach instead of bag-of-words approach for representing documents. We showed that our approach can provide precise and flexible results. It is also suggested that the outcome of predication – predication similarity can be used for question answering and knowledgebase exploration purposes. With the suggested improvements in the near future, we believe that the proposed approach can make a significant real world impact by a use case implementation for document retrieval.

## Acknowledgement

This research was supported in part by an appointment to the NLM Research Participation Program. This program is administered by the Oak Ridge Institute for Science and Education through an interagency agreement between the U.S. Department of Energy and the National Library of Medicine.